\documentclass[final,3p,times,twocolumn]{elsarticle}
\biboptions{sort&compress}

\usepackage{amssymb,mathrsfs}
\usepackage{amsthm,amsmath}
\usepackage{bm}
\usepackage{multirow,booktabs}
\usepackage{slashed}
\usepackage{threeparttable}

\usepackage[
bookmarks=true,colorlinks,%
citecolor=blue,linkcolor=blue,
breaklinks=true]{hyperref}

\begin{document}
	
	
	\hyphenpenalty=6000
	\tolerance=3000
	
	\begin{frontmatter}
		
		
		
        \title{Neutron Magic Numbers in $sd$ Shell from Nuclear Charge Radii within Neutron-Proton Correction around the Fermi Surface}
		
		
		\author[GXNU,GXNPT] {Yu-Ting Rong  \corref{cor1}}
        \ead{rongyuting@gxnu.edu.cn}
        \cortext[cor1]{Corresponding author.}
        
        \author[GXNU,GXNPT] {Ping-Mo Liu }
        \author[GXNU,GXNPT] {Dan Yang }
        
        \author[NXU,BJNU] {Rong An \corref{cor1}}
        \ead{rongan@nxu.edu.cn}

		\affiliation[GXNU]{%
		    organization={Department of Physics, Guangxi Normal University},%
		    city={Guilin},%
		    postcode={541004},%
		    country={China}}	
	    
	    \affiliation[GXNPT]{%
	    	organization={Guangxi Key Laboratory of Nuclear Physics and Technology, Guangxi Normal University},%
	    	city={Guilin},%
	    	postcode={541004},%
	    	country={China}}
		
		\affiliation[NXU]{%
			organization={School of Physics, Ningxia University},%
			city={Yinchuan},%
			postcode={750021},
			country={China}}
		
		\affiliation[BJNU]{%
			organization={Key Laboratory of Beam Technology of Ministry of Education, School of Physics and Astronomy, Beijing Normal University},%
			city={Beijing},%
			postcode={100875},%
			country={China}}

		\begin{abstract}
			Charge radii are sensitive indicators to identify the nuclear structure phenomena throughout the whole nuclide chart. In particular, the shrunken trend of changes of charge radii along a long isotopic chain is intimately associated with the shell quenching effect. In this work, the systematic evolution of charge radii along the proton numbers $Z=8$, 10, 12, 14, 18 isotopes is investigated by a relativistic Hartree Bogoliubov model. A ansatz about neutron-proton correlation around Fermi surface is considered for describing the abnormal behavior of nuclear charge radii. Our results show that the neutron-proton pairing corrections around the Fermi surface lead to a sudden strengthening of the charge radii of these isotopic chains at $N=8$, 20 and 28, reflecting the fact that this correction enhances the shell closure across $N=8$, 20 and 28. The reproduction of the $N=14$ charge radius in the Mg isotopes is affected by the way in which pairing correlations are handled, with BCS theory overestimating the shell effect of $N=14$, and the Bogoliubov quasiparticle transformation suggests a stronger pairing correlation near the proton Fermi surface, which is more consistent with experimental results. An analysis of the deviations from the theoretical and available experimental data for the charge radii of the 24 selected even-even nuclei shows that the neutron-proton pairing correction around the Fermi surface has an improved effect on the calculation of the charge radii using the meson-exchange effective interactions, but it does not help to significantly improve the results calculated by the density-dependent effective interactions.
	
		\end{abstract}

		
	\end{frontmatter}
	
\section{Introduction}\label{sec1}

Shell structure is a distinctive feature of nuclear many-body systems. Such structure is characterized by the existence of proton and neutron magic numbers. The nuclear magic numbers reveal the chemical stability and internal intrinsic structure of the atomic nucleus, which is of profound significance to the understanding of nuclear physics, element formation, and practical applications. 
Magic nuclei can be manifested by various phenomena, such as sudden increases in binding energy~\cite{Wang2021_ChinPhysC45-030003}, high  excited energies, low quadrupole transition probabilities~\cite{Pritychenko2016_ADNDT107-1}, localized abrupt changes in charge radius~\cite{Angeli2013_ADNDT99-69,Angeli2016_JPCS724-012032} and proton radius~\cite{Bagchi2019_PLB790-251,Kaur2022_PRL129-142502}, and reduced neutron/proton capture cross sections~\cite{DiazFernandez2018_PRC97-024311} compared to the neighboring nuclei. Modern radioactive beam facilities further extend the study of extreme nuclides, potentially revealing new magic numbers or revising traditional shell models~\cite{Sorlin2008_PPNP61-602,Yang2023_PPNP129-104005,Ye2025_NRP7-21}.

The systematic evolution of the bulk properties from oxygen to argon  isotopic chains embraces rich information about the shell structure, especially the disappearance of traditional magic numbers and the emergence of new ones. 
Along the oxygen isotopes, 
the highly excited energies and low $B(E2)$ value of the $2^+_1$ state~\cite{Thirolf2000_PLB485-16,Stanoiu2004_PRC69-034312}, small quadrupole transition parameter $\beta_2$~\cite{Becheva2006_PRL96-012501}, reduced proton radius~\cite{Kaur2022_PRL129-142502}, 
large inclusive cross section, and wide momentum distribution from quasifree ($p,pN$) scattering~\cite{DiazFernandez2018_PRC97-024311}, provide strong evidences for the existence of the $N=14$ subshell closure in $^{22}$O.
Also, a large shell gap is evidently identified for $N=16$ in $^{24}$O from various measurements~\cite{Ozawa2000_PRL84-5493,Kanungo2002_PLB528-58,Otsuka2001_PRL87-082502,Stanoiu2004_PRC69-034312,Hoffman2009_PLB672-17,Kanungo2009_PRL102-152501,Tshoo2012_PRL109-022501,Kaur2022_PRL129-142502}, manifesting that $^{24}$O is a doubly magic nucleus. Masses, charge radii and Coulomb excitation measurements of neutron-rich Ne, Na, and Mg isotopes~\cite{Chaudhuri2013_PRC88-054317,Marinova2011_PRC84-034313,Yordanov2012_PRL108-042504,Angeli2016_JPCS724-012032} have suggested the breakdown of the traditional magic number $N=20$~\cite{Thibault1975_PRC12-644,Motobayashi1995_PLB346-9,Pritychenko1999_PLB461-322,Yanagisawa2003_PLB566-84}.
Higher in mass, the observation of a low-lying $2^+_1$ state in $^{42}$Si provides transparent evidence for the collapse of the $N=28$ shell
closure~\cite{Bastin2007_PRL99-022503}.

Plenty of nuclear structure models are undertaken to unveil the underlying mechanisms of the emergence of new magicity and the collapse of shell quenching phenomena.
The results obtained by the shell model have suggested that the appearance of new magic numbers $N=14$ and $N=16$ is attributed to the strong neutron-proton tensor interaction~\cite{Otsuka2001_PRL87-082502,Otsuka2005_PRL95-232502,Bagchi2019_PLB790-251,Kaur2022_PRL129-142502}. Ab initio calculations with a novel version of two- and three-nucleon forces lead to considerable improvement in simultaneously describing the binding energies, charge and matter radii for stable O isotopes, but the deficiencies for the most neutron-rich systems is encountered~\cite{Lapoux2016_PRL117-052501}.
The collapse of the $N=20$ shell closure is attributed to
populating the neutron $pf$ shell in the presence of $sd$
orbitals at a considerable prolate deformation, which is known as ‘‘island of inversion’’~\cite{Poves1987_PLB184-311,Warburton1990_PRC41-1147}. The coupled-cluster method based on nucleon-nucleon and three-nucleon
potentials qualitatively reproduced the evolutionary trends in
charge radii and the neutron number $N = 14$ of Ne and Mg isotopes after taking angular momentum projection into account, but isotope shifts still challenge~\cite{Novario2020_PRC102-051303}.
{In mean-field theories, the ground state of $^{32}$Mg is spherical, and become deformed after increase
	by 20\% of the spin-orbit strength based on Skyrme force  SLy4~\cite{Gaidarov2014_PRC89-064301}; A deformed ground state can be achieved by adjusting the neutron and proton pairing gaps, but the magic number $N = 20$ still exists for the ground state of $^{32}$Mg in the relativistic mean-field (RMF) theory~\cite{Ren1996_PLB380-241}. Angular momentum projection approach based on HFB model~\cite{RodriguezGuzman2000_PLB474-15,Borrajo2017_PLB764-328} and RMF model~\cite{Yao2009_PRC79-044312} changes the spherical mean-field ground state of $^{32}$Mg to deformed one with $\beta_2$ close to the measured value. Similar picture is also obtained from the projected shell model~\cite{Dong2013_PRC88-024328}.}

As mentioned above, the shrunken trend of changes of the charge radii along a isotopic family is one of the signatures to identify the shell closure effect. The charge radii are influenced by various mechanisms, such as pairing correlation~\cite{Reinhard2017_PRC95-064328,Perera2021_PRC104-064313,Miller2019_PLB793-360,Souza2020_PRC101-065202,Cosyn2021_PLB820-136526},
deformation~\cite{Lalazissis1996_NPA597-35,An2023_CTP75-035301}, cluster structure~\cite{Mueller2007_PRL99-252501,Geithner2008_PRL101-252502}, 
shell evolution~\cite{Barzakh2018_PRC97-014322,Nakada2019_PRC100-044310,DayGoodacre2021_PRL126-032502}, 
and center-of-mass correlation~\cite{Long2004_PRC69-034319,Reinhard2021_PRC103-054310,Rong2023_PRC108-054314}. 
The modified RMF plus
BCS equation ansatz (RMF(BCS)* model), in which neutron-proton pairing correlations around Fermi surface have been incorporated into the charge radii formula~\cite{An2020_PRC102-024307, An2022_CPC46-054101}, describes the odd-even staggerings and the inverted parabolic-like behavior well.
This method gives a good description of the charge radii for most of O, Ne and Mg isotopes. However, it predicts the existence of odd-even staggerings in the charge radii of the O isotopic chain with a sudden increase at $N=14$; underestimates the charge radii of the nuclei for $N=14$ and $N>18$ nuclei in the Mg isotopes; and overestimates the charge radius of $^{18}$Ne ($N=8$).
Considering the pairing correlation by the BCS approximation is not suitable for nuclei far away from the $\beta$-stability line~\cite{Dobaczewski1984_NPA422-103,Dobaczewski1996_PRC53-2809},  and the magic numbers 14, 16, and 20 in $sd$ shell are appeared in neutron-rich region, and even at the drip line.
Therefore, it is necessary to further study the charge radii with the same ansatz by treating the pairing correlation with a more suitable method.

Tackling pairing correlation by the Bogoliubov transformation is considered to be better than the BCS method for nuclei far away from $\beta$-stability line. In this consideration, we recently introduce the neutron-proton pairing correlation extracted from the quasiparticle states around Fermi surface to the multidimensionally-constrained relativistic Hartree-Bogoliubov (MDC-RHB) model~\cite{Yang2024_PRC110-064314}. The charge radii of Ca and Ni isotopes, and $N = 28$, 30, 32, and 34 isotones are well reproduced. Therefore, in this work, we can revisit the problems encountered in the RMF(BCS)* approach and the shell closure effect around the neutron numbers $N=14, 16$ and 20 from the aspects of nuclear charge radii, manifest the relation between the shell structure in these isotopes and the neutron-proton pairing.
This work is organized as follows. 
In Sec.~\ref{sec:model}, 
we briefly introduce the MDC-RHB model and the neutron-proton pairing correlation extracted from the quasiparticle states around Fermi surface. 
In Sec.~\ref{sec:results}, the charge radii for $Z=8$, $10$, $12$, $14$,  and $18$ isotopes are
investigated with this method, and the shell structure phenomena around $N=14$, $16$ and $20$ are discussed as well.  A summary is given in Sec.~\ref{sec4}.

\section{Theoretical framework}\label{sec:model}

In the MDC-RHB model, the RHB equation in coordinate space can be recalled as follows~\cite{Ring1996_PPNP37-193,Kucharek1991_ZPA339-23}:
\begin{equation}
	\label{eq:rhb}
	\int d^{3}\bm{r}^{\prime}
	\left( \begin{array}{cc} \bm{h}-\lambda  &  \Delta                      \\
		-\Delta^{*}   & -\bm{h}+\lambda \end{array}
	\right)
	\left( \begin{array}{c} U_{k} \\ V_{k} \end{array} \right)
	= E_{k}
	\left( \begin{array}{c} U_{k} \\ V_{k} \end{array} \right),
\end{equation}
where $\lambda$ is the Fermi energy, $\Delta$ is the pairing tensor field, $E_k$ is the quasiparticle energy, and $(U_k(\bm{r}),V_k(\bm{r}))^T$ is the quasiparticle wave function.
$\bm{h}$ is the single-particle Hamiltonian which can be written as follows:
\begin{equation}
	\bm{h} = \bm{\alpha} \cdot \bm{p}  +
	\beta \left[ M + S(\bm{r}) \right]+ V(\bm{r})+\Sigma_R(r),
\end{equation}
where $M$ is the mass of nucleon, $S(\bm{r})$, $V(\bm{r})$, and $\Sigma_R(r)$ are the scalar, vector, and rearrangement potentials, respectively.
For meson-exchange interactions,
\begin{equation}
	\begin{aligned}
		&S(\bm{r})=g_\sigma \sigma, \\
		&V(\bm{r})=g_\omega \omega_0+g_\rho \rho_0 \cdot \tau_3+e\dfrac{1-\tau_3}{2}A_0, \\
		&\Sigma_R(\bm{r})=\dfrac{\partial g_\sigma}{\partial \rho_V} \rho_S \sigma + \dfrac{\partial g_\omega}{\partial \rho_V} \rho_V \omega_0 + \dfrac{\partial g_\rho}{\partial \rho_V} \rho_V\tau_3 \rho_0,
	\end{aligned}  
\end{equation} 
where $g_\sigma$, $g_\omega$, and $g_\rho$ are coupling constants of $\sigma$, $\omega_0$, and $\rho_0$ meson fields, $A_0$ is the time-like component of the Coulomb field mediated by photons naturally, $e$ is the charge unit for protons. The quantities of $\tau_3=1$ and $-1$ are used to distinguish the corresponding components of neutron and proton. $\rho_S$ and $\rho_V$ are the isoscalar and isovector densities, respectively.
For point-coupling interactions,
\begin{equation}
	\begin{aligned}
		S(\bm{r})=&\alpha_S\rho_S+\alpha_{TS}\rho_{TS}\tau_3 + \beta_S	\rho_S^2 +\gamma_S\rho_S^3 \\
		&+\delta_S\Delta \rho_S +\delta_{TS}\Delta \rho_{TS}\tau_3, \\
		V(\bm{r})=&\alpha_V\rho_V+\alpha_{TV}\rho_V\tau_3 +\gamma_V\rho_V^3 \\
		&+\delta_V\Delta\rho_V+\delta_{TV}\Delta \rho_{TV}\tau_3 +e\dfrac{1-\tau_3}{2}A_0, \\
		\Sigma_R(\bm{r})=&\dfrac{1}{2} \dfrac{\partial \alpha_S}{\partial \rho_V} \rho_S^2 
		+ \dfrac{1}{2} \dfrac{\partial \alpha_V}{\partial \rho_V} \rho_V^2 +\dfrac{1}{2} \dfrac{\partial \alpha_{TV}}{\partial \rho_{V}} \rho_{TV}^2,
	\end{aligned}  
\end{equation}
where $\alpha_S$, $\alpha_V$, $\alpha_{TS}$, $\alpha_{TV}$, $\beta_S$, $\gamma_S$, $\gamma_V$, $\delta_S$, $\delta_V$, $\delta_{TS}$, and $\delta_{TV}$ are coupling constants for different channels, $\rho_{TS}$ and $\rho_{TV}$ are time-like components of isoscalar current and time-like components of isovector current, respectively.
The pairing field $\Delta$ is calculated by the effective pairing interaction $V$ and the pairing tensor $\kappa$ as follows:
\begin{equation}
	\begin{aligned}
		&\Delta(\bm{r}_{1}\sigma_{1},\bm{r}_{2}\sigma_{2}) 
		= 
		\int d^{3}\bm{r}_{1}^{\prime} d^{3}\bm{r}_{2}^{\prime} 
		\sum_{\sigma_{1}^{\prime}\sigma_{2}^{\prime}}  \\
		&	V(\bm{r}_{1}         \sigma_{1},          \bm{r}_{2}         \sigma_{2},
		\bm{r}_{1}^{\prime}\sigma_{1}^{\prime}, \bm{r}_{2}^{\prime}\sigma_{2}^{\prime}) 
		\kappa 
		(\bm{r}_{1}^{\prime}\sigma_{1}^{\prime}, 
		\bm{r}_{2}^{\prime}\sigma_{2}^{\prime}).
	\end{aligned}
\end{equation}
In this work, a separable pairing force of finite
range with pairing strength $G = 728$ MeV fm$^3$ and effective
range of the pairing force $a = 0.644$ fm are adopted~\cite{Tian2009_PLB676-44}.

The modified root-mean-square (rms) charge radii $r_{\rm ch}$ is recalled as follows ~\cite{An2024_PRC109-064302}:
\begin{equation}\label{eq:r_ch}
	r_{\rm ch}^2=\langle r_p^2 \rangle +0.7056~{\rm fm}^2+\dfrac{a_0}{\sqrt{A}}\Delta D~{\rm fm}^2 +\dfrac{\delta}{\sqrt{A}}~{\rm fm}^2.
\end{equation}
where the first term represents the charge distributions of point-like protons, and the second term is incorporated due to the finite
size of protons, $A$ is the mass number, and ${a_0}$ is a normalization constant. The term $\Delta D=|D_n-D_p|$ is employed to measure the neutron-proton correlations around Fermi surface, which reads: 
\begin{equation}\label{eq:Dnp}
	D_{n,p}=\sum_{k>0} u_k^{n,p} \upsilon_k^{n,p},
\end{equation}
where $u_k^{n,p}$ is the occupation probability of the $k$th quasiparticle orbital for neutron or proton, and $\upsilon_k^2={1-u_k^2}$. In practice, the quasiparticle levels that satisfy $|E_k-\lambda|<20$ MeV are summed up in Eq.~(\ref{eq:Dnp}). The
values of $a_0=$0.561 and $\delta=$0.355(0.000) for odd-odd (even-even, odd-even, and even-odd) nuclei are the same as those shown in Ref.~\cite{An2024_PRC109-064302}. 

In the MDC-RHB model, one can allow for mutipole moments under $V_4$ symmetry~\cite{Lu2014_PRC89-014323,Zhou2016_PS91-063008}. In this work, we restrict the calculations into both axial and reflection symmetries, namely, only quadrupole deformation $\beta_{20}$ is considered. The effective interactions PK1~\cite{Long2004_PRC69-034319}, NL3~\cite{Lalazissis1997_PRC55-540}, DD-ME2~\cite{Lalazissis2005_PRC71-024312}, and DD-PC1~\cite{Niksic2008_PRC78-034318} are used in our calculations to avoid parameter-dependent results.

\section{Results and discussion}\label{sec:results}

\begin{figure}[htbp]
	\centering
	\includegraphics[width=0.48\textwidth]{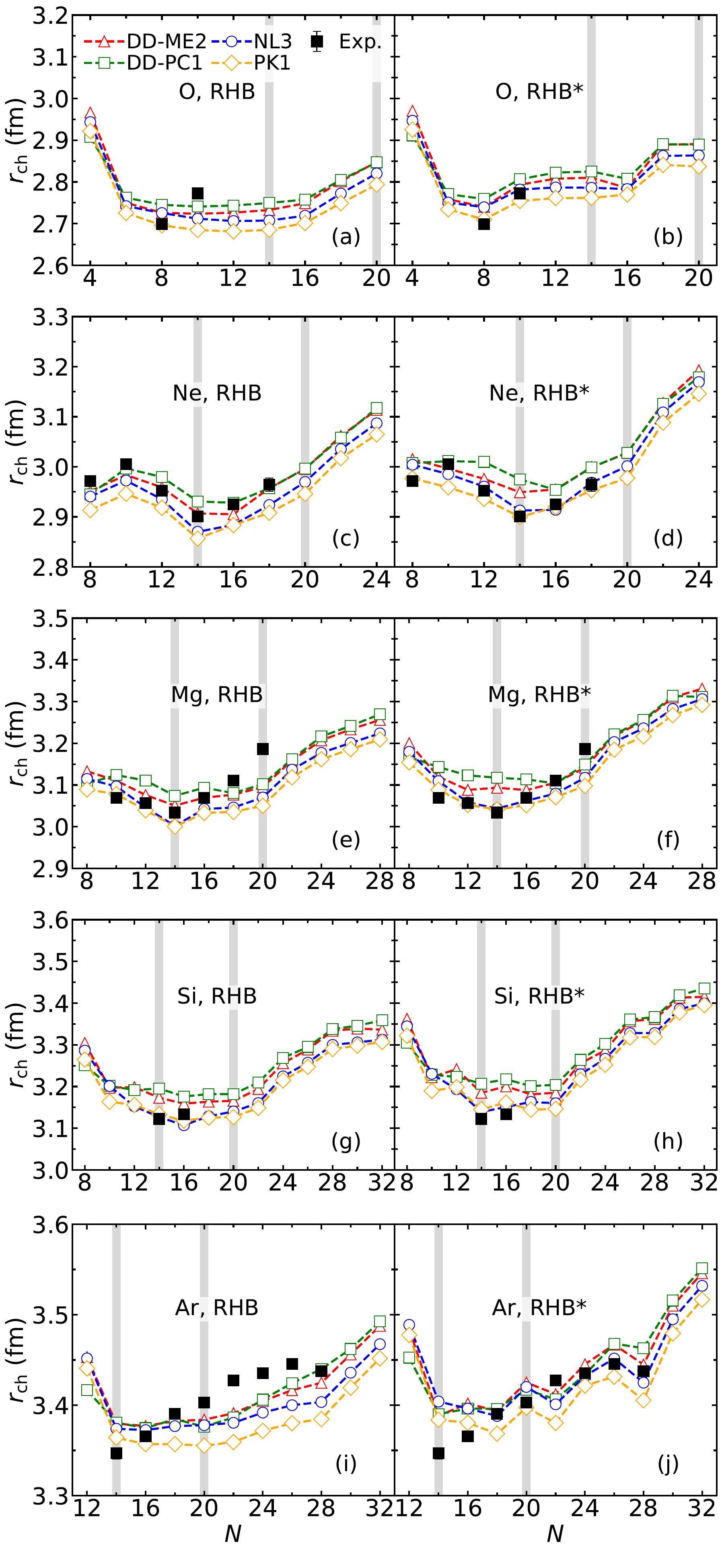}
	\caption{(Color online) Charge radii of O, Ne, Mg, Si, and Ar isotopes calculated by the RHB (left) and RHB* (right) models with the DD-ME2, NL3, DD-PC1, and PK1 effective interactions. The corresponding experimental data are taken from Refs.~\cite{Angeli2013_ADNDT99-69,Yordanov2012_PRL108-042504} (solid square). The gray band represents the neutron numbers $N = 14$ and $20$.}
	\label{fig:radii_RHB}
\end{figure}

As mentioned in the introduction part, the $N=14$ and 16 shell quenching effects are observed in O, N, Ne, and Mg isotopes. 
To investigate these characteristic  magicities, the systematic evolution of charge radii along O, Ne, Mg, Si, and Ar isotopic chains are calculated with the MDC-RHB model.
For the sake of discussion, the results obtained with and without the neutron-proton correlations around Fermi surface are marked by the RHB* and RHB models, respectively.

The charge radii {along O, Ne, Mg, Si, and Ar isotopic chains are depicted with the DD-ME2, NL3, DD-PC1, and PK1 effective interactions as shown in Fig.~\ref{fig:radii_RHB}.} 
The RHB calculations yield charge radii for $^{16-22}$O within the same order of magnitude as shown in Fig.~\ref{fig:radii_RHB}(a). Among of them, the PK1 force provides the most accurate estimate for $^{16}$O. However, the charge radius for $^{18}$O is underestimated by all of the four forces. 
As shown in Fig.~\ref{fig:radii_RHB}(b), the systematic trend of changes of charge radii is depressed at $N = 8$ and $16$ with the RHB* model. This implies that the shell quenching at $N=8$ and $16$ are enhanced after the implementation of the neutron-proton correction around the Fermi surface.
The correction term has the effect of increasing the charge radius of $^{18}$O, therefore improving the description of its charge radius provided by {the PK1 and NL3 sets}. The double-magic property in $^{16}$O {makes largely free of surface proton-neutron correlations}, thereby maintaining the charge radius as it was prior to the correction.

The charge radii for Ne isotopes are shown in Fig.~\ref{fig:radii_RHB}(c) and (d).
The results obtained from the RHB calculations are in alignment with the trend of the experimental results concerning isotopic evolution, wherein the density-dependent forms of the effective interactions DD-ME2 and DD-PC1 are more proximate to the experimental values, and the meson-exchange effective interactions {(NL3 and PK1)} underestimate the experimental values a little. 
The kinks in the charge radii {can be observed at the $N=8$ and $N=14$ shell closures}.
After the consideration of the neutron-proton surface pairing correction, the charge radii are increased and those calculated with meson-exchange effective interactions are more in line with the experimental data. The charge radius is smallest calculated by DD-ME2, NL3, and PK1 at $N=14$. However, DD-PC1 yielded the smallest value of charge radius at $N=16$.
Furthermore, with increasing neutron numbers, the RHB model predicts that there will be no abrupt change in charge radii across the neutron number $N=20$.
In contrast, the RHB* model anticipates that the surface pairing of protons and neutrons will result in a sudden increase in the charge radius of $^{32}$Ne, thereby manifesting the magicity of $N=20$. It is important to note that the RHB* model exhibits suboptimal performance on the proton side, leading to an overestimation of the charge radius for $^{18}$Ne.

For Mg isotopic chain, as illustrated in Fig.~\ref{fig:radii_RHB}(e), the charge radii of proton-rich isotopes are in close agreement with the experimental values in the RHB calculations. However, the charge radii of the neutron-rich isotopes are underestimated, exhibiting kinks at $N=14$ and 20. As illustrated in Fig.~\ref{fig:radii_RHB}(f), the Fermi-surface correlation between protons and neutrons leads to an enhancement in the charge radii of isotopes with a neutron number greater than 14. This phenomenon results in RHB* calculations that exhibit greater concordance with experimental measurements compared to RHB calculations. 
Therefore, within the neutron-rich side of Mg isotopes, this neutron-proton pairing surface vibration may serve as the underlying mechanism responsible for the observed change in the charge radii. A comparison of Ref.~\cite{An2020_PRC102-024307} reveals that the charge radius depression of $N=14$ in that work is too large, given that the same charge radius correction method was employed. In contrast, the charge radius obtained by the RHB* model using NL3 and PK1 effective interactions are 3.0434 fm and 3.0415 fm, respectively, which are very close to the experimental value of 3.0340(26) fm~\cite{Yordanov2012_PRL108-042504}. {This finding indicates that the characterization of the $N=14$ magicity exerts a significant influence on the treatment of the correlations}. Therefore, it is necessary to investigate the form of the pairing force within the same mean-field Hamiltonian and to thoroughly analyze the microscopic mechanism by which the $N=14$ charge radius produces the aforementioned results.

The charge radii of the Si isotopes are displayed in Fig.~\ref{fig:radii_RHB}(g) and (h). It can be seen that the charge radius of $^{28}$Si ($N=14$) calculated by RHB model is larger than that of $^{30}$Si ($N=16$), which is opposite to the experimental data. {Considering the neutron-proton correlation around Fermi surfaces}, the charge radii of these two nuclei are in good agreement with the experimental data. Among them, the theoretical values given by PK1 and NL3 effective interactions are comparable to the experimental values. For $N=20$ and 28, the charge radii appear more pronounced kinks after considering the correction term, indicating the enhancement of shell closures at $N=20$ and 28.

Figures~\ref{fig:radii_RHB}(i) and (j) show the charge radii of Ar isotopes calculated by the RHB and RHB* models, respectively. The charge radii calculated by the RHB model show a “parabolic” form as a function of the neutron numbers, which exhibits larger deviation from the experimental results. The charge radii calculated by the RHB* model are basically in good agreement with the experimental results at the neutron-rich side, except for $N=22$ where the charge radius is somewhat underestimated.
On the proton-rich side, the charge radius of $N=14$ isotope deviates heavily from the experimental value after the correction term is taken into account.

\begin{figure}[htbp!]
	\centering
	\includegraphics[width=0.35\textwidth]{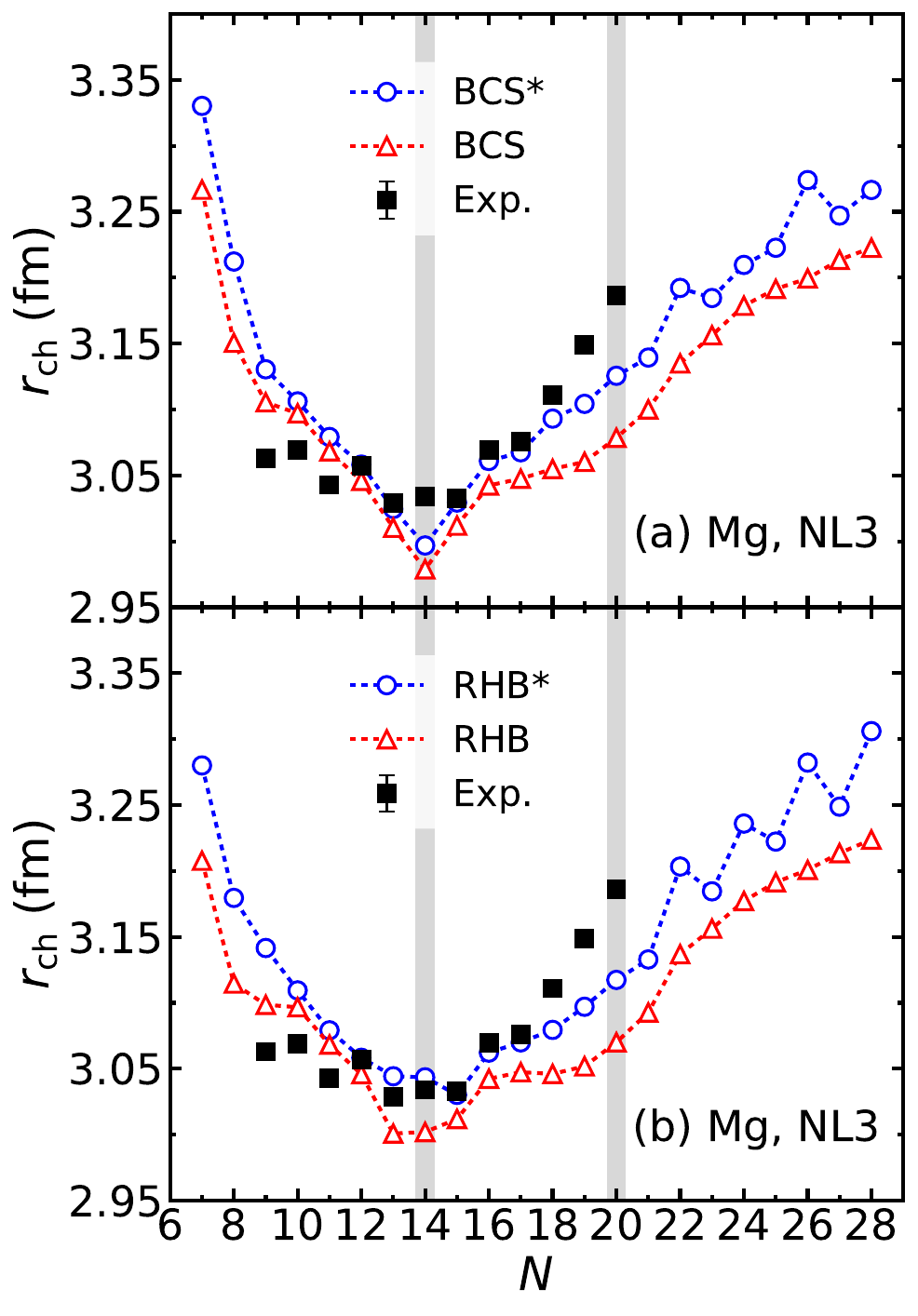}
	\caption{(Color online) Charge radii for Mg isotopes with (a) BCS and (b) RHB treatment on pairing correction. The results are calculated with NL3 effective interaction based on the same single-particle Hamiltonian. The corresponding experimental data are taken from Ref.~\cite{Yordanov2012_PRL108-042504} (solid square). The gray band represents the neutron numbers $N = 14$ and $20$.}
	\label{fig:Mg_radii}
\end{figure}

\begin{figure*}[htbp!]
	\centering
	\includegraphics[width=0.9\textwidth]{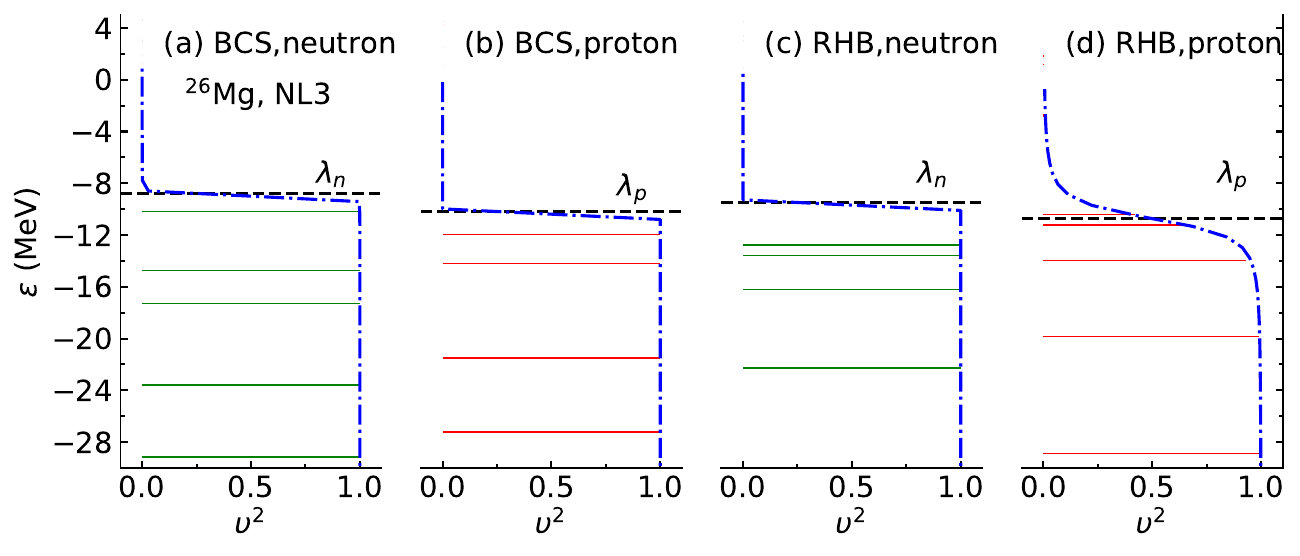}
	\caption{(Color online) Single-particle levels of ground state of $^{26}$Mg as a function of occupation probability $\upsilon^2$ calculated with RMF+BCS and RHB models. The NL3 effective interaction is adopted. The black dashed lines are the Fermi surfaces. The blue
		dashed-dotted line corresponds to the BCS formula with an average pairing gap. }
	\label{fig:Mg26_SPL_NL3}
\end{figure*}

As mentioned above, the local variations of nuclear charge radii can be influenced by the treatment of the pairing correlations. Thus, we show in Fig.~\ref{fig:Mg_radii} the calculations of the charge radii along Mg isotopic chain given by the BCS and Bogoliubov quasiparticle transformations, respectively, with the same single-particle Hamiltonian $\bm{h}$.
Given the concurrence of the corrected charge radius with the experimental values, this calculation employs the NL3 effective interaction. 
As illustrated in Fig.~\ref{fig:Mg_radii}(a), the charge radius determined by the BCS pairing correlation exhibits a pronounced trough at $^{26}$Mg, both prior to and following neutron-proton surface correction. This observation aligns with the findings reported in the literature~\cite{An2020_PRC102-024307}, suggesting a strong magicity for $N=14$. 
In Fig.~\ref{fig:Mg_radii}(b), the application of the Bogoliubov quasiparticle transformation to address the pair correlation reveals that the charge radius of $^{24-29}$Mg, after neutron-proton correction, exhibits a strong concurrence with the experimental data. That is, this correction facilitates the reproduction of the charge radius at the neutron number $N=14$.

{In Fig.~\ref{fig:Mg26_SPL_NL3}, we analyze the depression of the charge radius at $N=14$ and the microscopic mechanism by which it is affected by the surface pairing corrections from the $^{26}$Mg single-particle energy levels.}
Figures~\ref{fig:Mg26_SPL_NL3}(a) and (b) are the neutron and proton single-particle energy levels calculated using the RMF+BCS method, respectively. It is evidently seen from these figures that the neutron and proton single-particle levels are predominantly occupied below the Fermi surface, the number of occupations above the Fermi surface is nearly zero. This finding indicates that the pairing gaps for both $N=14$ and $Z=12$ are relatively small. This, in turn, suggests that the nuclei are more stable and have small charge radii. According to Eq.~(\ref{eq:r_ch}), for a given nucleus, the neutron-proton pairing correlation around the Fermi surface is determined by $\Delta D$. The latter is, in turn, determined by the non-integer occupation numbers of the protons and neutrons. The calculated results with the BCS method indicate that the particles are predominantly integer occupied, resulting in a small $\Delta D$ and an insignificant correction effect. The single-particle energy levels obtained by using the RHB model (see Fig.~\ref{fig:Mg26_SPL_NL3}(c) and (d)) demonstrate that the neutron single-particle energy levels are consistent with the results obtained by the BCS method. However, there are obviously fractional occupations in several single-particle energy levels near the proton Fermi surface obtained by the RHB calculations. In essence, the proton pairing correlation enhancement results in a charge radius that is calculated by the RHB method to be slightly larger than the radius calculated by the BCS method. Concurrently, the augmentation of $D_p$ in Eq.~(\ref{eq:r_ch}) and the escalation of the correction term for the neutron-proton surface pairing correlation result in the $^{26}$Mg charge radius calculated by RHB* attaining a magnitude that is substantially greater than that calculated by RHB model. This outcome is more congruent with experimental measurements.

 The observed absence of the $N=20$ shell closure in Mg isotopes is typically attributed to the presence of inversion islands. The inversion of energy levels results in a substantial deformation in the ground state of $^{32}$Mg, leading to the dissolution of the $N=20$ shell closure. However, the RHB model fails to replicate this phenomenon, as all four sets of parameters yield spherical ground states. 
	Beyond mean-field calculations yielded reasonably large deformation in the ground states of $^{32}$Mg~\cite{RodriguezGuzman2000_PLB474-15,Borrajo2017_PLB764-328,Yao2009_PRC79-044312}, could able to address the observed discrepancy between the calculated and measured charge radii. 
	In the present study, we demonstrate that RHB* calculation also yield a more accurate charge radius for $^{32}$Mg in comparison with RHB calculation, especially calculated with the density-dependent effective interaction DD-PC1 and DD-ME2. A beyond mean-field calculation based on the present framework can be seen from Refs.~\cite{Wang2022_CTP74-015303,Rong2023_PLB840-137896}. Further study in this direction is needed.

\begin{figure}[htbp!]
	\centering
	\includegraphics[width=0.38\textwidth]{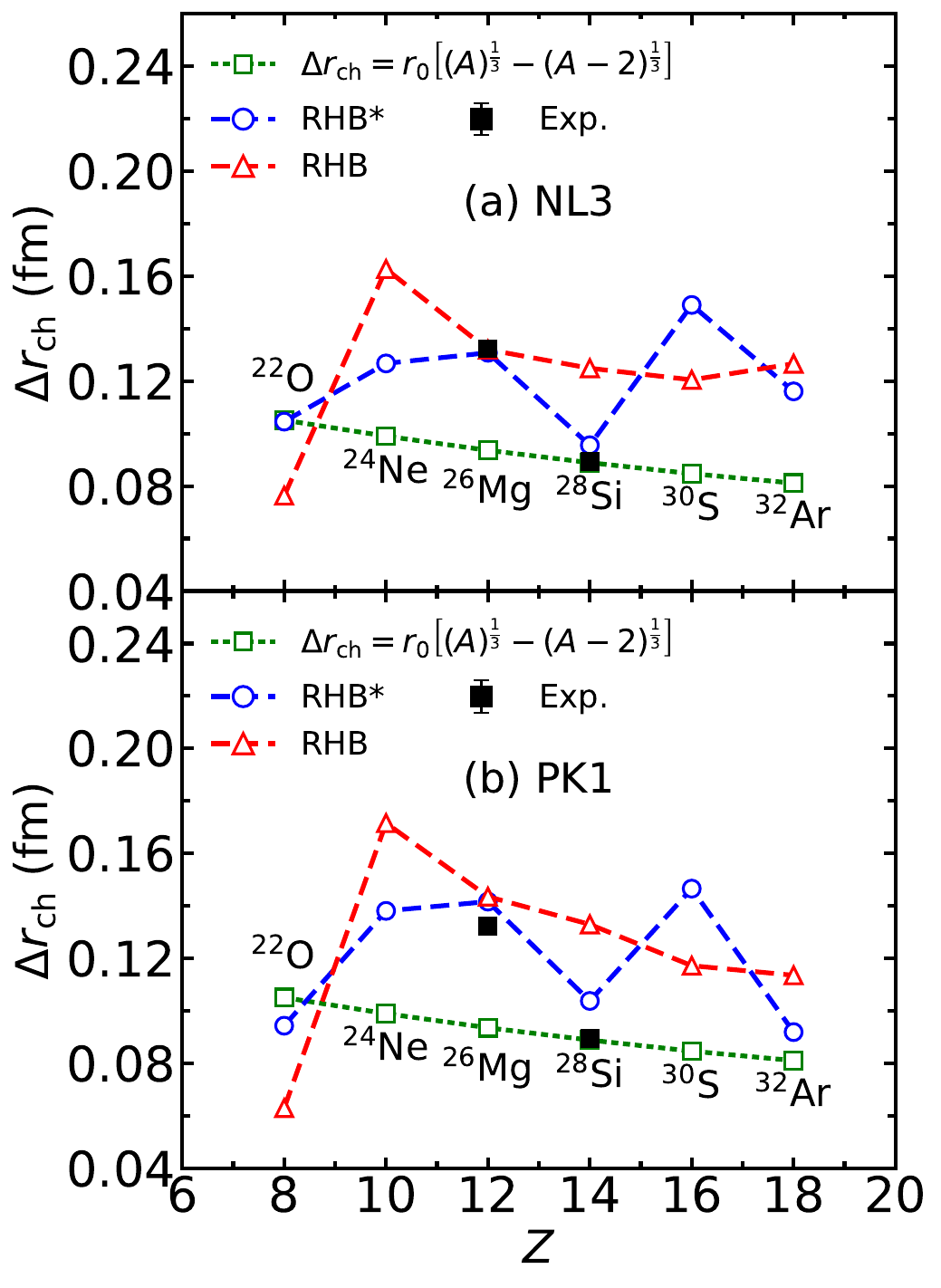}
	\caption{(Color online) $\Delta r_{\rm ch}$ as a function of proton number for $N=14$ isotones calculated with (a) NL3 and (b) PK1 effective interactions. The corresponding experimental data are taken from Refs.~\cite{Angeli2013_ADNDT99-69,Yordanov2012_PRL108-042504} (solid square). The empirical values with $r_{\rm ch}=r_0 A^{1/3}$ ($r_0=1.2$ fm) are shown as guideline.}
	\label{fig:DelRch_RHB}
\end{figure}

To facilitate the understanding of the $N = 14$ magicity, the $\Delta r_{\rm ch}$, defined as 
\begin{equation}
	\Delta r_{\rm ch}(Z,N)= r_{\rm ch}(Z,N)-r_{\rm ch}(Z-2,N),
\end{equation}
along $N = 14$ isotones are presented in Fig.~\ref{fig:DelRch_RHB}. 
As can be seen from the figure, the calculations given by NL3 and PK1 are basically the same. As shown by the green line in the figure, the empirical formula $1.2A^{1/3}$ fm is only related to the mass number and cannot give the shell structure. The results of the RHB calculations show a very small $\Delta r_{\rm ch}(Z)$ in the charge radius for $Z=8$, implying that the  magicity of $N=14$ is very strong in $^{22}$O, which is in agreement with the measurements of the proton radius of $^{22}$O~\cite{Kaur2022_PRL129-142502}. After considering the correction comes from neutron-proton pairing, the shell closure of $N=14$ in $^{22}$O is weakened, and that in $^{28}$Si is enhanced. Combined with the experimental values for $^{26}$Mg, the RHB* model gives a better picture of the currently available experimental results than the RHB model.

Finally, the average deviation $\bar{\chi}^2$ and the root mean square deviation $\Delta$ (defined as follows)
\begin{equation}\label{eq:deviation}
	\begin{aligned}
		&\bar{\chi}^2=\dfrac{1}{N} \sum_{i}^N\left( \dfrac{ r_{{\rm ch},i}^{\rm exp.} - r_{{\rm ch},i}^{\rm cal.} }   {\Delta r_{{\rm ch},i}^{\rm exp.} }\right)^2,\\
		&\Delta=\sqrt{ \dfrac{1}{N} \sum_{i}^N \left( r_{{\rm ch},i}^{\rm exp.} - r_{{\rm ch},i}^{\rm cal.} \right)^2 } ,
	\end{aligned}
\end{equation}
between the 24 experimental and calculated charge radii are listed in Table~\ref{tab:deviation}. The $\Delta r_{{\rm ch},i}^{\rm exp.}$ in Eq.~(\ref{eq:deviation}) is the experimental uncertainty of the corresponding measured charge radius $r_{{\rm ch},i}^{\rm exp.}$ for the $i$th
nucleus. 
Calculated with DD-ME2, NL3, and PK1, the $\bar{\chi}^2$ of the RHB* model is smaller than that of the RHB model. Conversely, the result calculated using the DD-PC1 effective interaction is reversed. When evaluated using the root-mean-square deviation $\Delta$, the calculated $\Delta$ of the meson-exchange effective interactions NL3 and PK1 decrease after correction, while those of the density-dependent effective interactions DD-ME2 and DD-PC1 increase slightly after correction. The aforementioned results indicate the necessity of incorporating neutron-proton pairing corrections around the Fermi surface in order to facilitate mean-field calculations of effective interactions based on meson exchange. This approach is imperative for the systematic description of the charge radius from the even-even O to Ar  nuclei. Density-dependent effective interactions are not required to account for this correction, as RHB calculations yield larger charge radii compared to meson-exchange effective interactions.

\begin{table}[htb!]
	\caption{The average deviation $\bar{\chi}^2$ and the root mean square deviation $\Delta$ between the 24 experimental and calculated charge radii of even-even nuclei from O to Ar isotopes.}\label{tab:deviation} 
	\doublerulesep 0.1pt \tabcolsep 6pt
	\begin{tabular}{crrrr}
		\hline
		\hline
		Force  &  \multicolumn{2}{c}{$\bar{\chi}^2$} & \multicolumn{2}{c}{$\Delta$}  \\
		\cmidrule(lrr){2-3} \cmidrule(lrr){4-5}
		&  RHB    &  RHB*  & RHB       &  RHB*\\
		\hline
		DD-ME2  &  648.80  &  479.61  &  0.0319  &  0.0361    \\
		DD-PC1  &  952.21  &  1083.08  &  0.0375  &  0.0482    \\
		NL3     &  990.43  &  427.34  &  0.0413  &  0.0266    \\
		PK1     &  1625.83  &  951.87  &  0.0719  &  0.0699    \\
		\hline         
		\hline	
	\end{tabular}
\end{table}

\section{Summary}\label{sec4}

	In this work, we systematically investigate the charge radii evolution across oxygen to argon isotopic chains using the RHB framework. A refined charge radius formula incorporating neutron-proton pairing correlations near the Fermi surface is considered for describing the abnormal behaviour. We analyse the emergence of new magic numbers at $N=14$ and 16, and the quenching of the traditional
	$N=20$ shell closure. 
	
	Notably, the inclusion of neutron-proton pairing correlations induces abrupt enhancements in the charge radii at 
	$N=8$, 20, and 28, demonstrating that such correlations amplify shell closure effects at these neutron numbers.
	The Mg isotopic chain provides critical insights into the interplay between the pairing treatments and shell structure. While conventional BCS theory artificially amplifies the $
	N=14$ shell effect, the Bogoliubov transformation approach predicts enhanced proton surface pairing correlations, yielding results in closer agreement with experimental charge radii. Further validation against experimental data for 24 even-even nuclei confirms that the neutron-proton pairing correction significantly improves the description for charge radius when using meson-exchange effective interactions, though its diminishes for density-dependent interactions. Future studies extend this framework to heavier isotopic chains is in progress. Also, the interaction-dependent performance of the neutron-proton pairing correction calls for a unified theoretical approach to reconcile the meson-exchange and density-dependent interactions.

\section*{Acnowledgements}

	This work has been supported by the National Natural Science Foundation of China (12205057), the Science and Technology Plan Project of Guangxi (Guike AD23026250), the Natural Science Foundation of Guangxi (2024JJB110017, 2023GXNSFAA026016), the Young Elite Scientists Sponsorship Program by GXAST (2025YESSGX159), the National Natural Science Foundation of China (12365016), and the Central Government Guides Local Scientific and Technological Development Fund Projects (Guike ZY22096024). R. A. was supported by the Open Project of Guangxi Key Laboratory of Nuclear Physics and Nuclear Technology, No. NLK2023-05, the Natural Science Foundation of Ningxia Province, China (No. 2024AAC03015), and the Key Laboratory of Beam Technology of Ministry of Education, China (No. BEAM2024G04). The results shown in this paper are obtained through the Guangxi Key Laboratory of Nuclear Physics and Technology High Performance Computing Platform.

	\bibliographystyle{elsarticle-num_20210104-sgzhou}
	\bibliography{ref.bib}
	
\end{document}